\documentclass[12pt]{article}
\begin{document}
\textwidth 9.0in
\textheight 9.0in
\topmargin -0.60in
\begin{center}
{\large \bf The Canonical Structure of the First Order\\
Einstein-Hilbert Action}\vspace{.5cm}\\
{D.G.C. McKeon\\
Department of Applied Mathematics\\
University of Western Ontario\\
London, ON \quad N6A 5B7\\
CANADA\vspace{.5cm}\\
and\vspace{.5cm}\\
Department of Mathematics and Computer Science\\
Algoma University\\
Sault Ste. Marie, ON\quad P6A 2G4\\
CANADA }
\end{center}
Email: dgmckeo2@uwo.ca
\begin{abstract}

The Dirac constraint formalism is used to analyze the first order form of the Einstein-Hilbert action
in $d > 2$ dimensions. Unlike previous treatments, this is done without eliminating fields at the outset by
solving equations of motion that are independent of time derivatives when they correspond to first class constraints.  As
anticipated by the way in which the affine connection transforms under a diffeomorphism, not only primary and
secondary but also tertiary first class constraints arise.  These leave $d(d-3)$ degrees of freedom in phase space.  The
gauge invariance of the action is discussed, with special attention being paid to the gauge generators of Henneaux, Teitelboim and Zanelli
and of Castellani.
\end{abstract}

\section{Introduction}

Einstein's formulation of general relativity leaves unanswered the question of which variables are most
appropriate to discuss the canonical structure of the theory.  The second order formalism in which the Einstein-Hilbert (EH)
action in $d$ dimensions
$$ S_d = \int d^dx \sqrt{-g} R\eqno(1)$$
is expressed in terms of the metric $g_{\mu\nu}$ has been treated in refs. [1-9]; the first order formalism with the
metric $g_{\mu\nu}$ and symmetric affine connection $\Gamma_{\mu\nu}^\lambda$ being independent was discussed for $d = 4$ in
refs. [10, 11] and for $d = 2$ in [12-17].

In this first order formalism, the Lagrangian can be taken to be [18-19]
$$
{\cal{L}}_d = h^{\mu\nu} \left[ G_{\mu\nu , \lambda}^\lambda + \frac{1}{d-1}  \,G_{\lambda\mu}^\lambda G_{\sigma\nu}^\sigma -
G_{\sigma\mu}^\lambda G_{\lambda\nu}^\sigma\right]\eqno(2)$$
where
$$ h^{\mu\nu} = \sqrt{-g}\, g^{\mu\nu}\,\,\,\left(det\,\, h^{\mu\nu} = -(-g)^{-1+d/2}\right)\eqno(3)$$
and [20]
$$
G_{\mu\nu}^\lambda = \Gamma_{\mu\nu}^\lambda -  \frac{1}{2} \left( \delta_\mu^\lambda \Gamma_{\rho\nu}^\rho +
\delta_\nu^\lambda \Gamma_{\rho\mu}^\rho\right).\eqno(4) $$

Only if $d > 2$ can the equations of motion for $G_{\mu\nu}^\lambda$ following from eq. (2) be solved for $G_{\mu\nu}^\lambda$
in terms of $h^{\mu\nu}$; if this solution is substituted back into eq. (2) the second order form of the EH action is recovered.\footnote[1]{From eq. (2),
if $d > 2$ then the equation of motion for $G_{\mu\nu}^\lambda$ yields $G_{\mu\nu}^\lambda = \frac{1}{2} h^{\lambda\rho} \left(
h_{\mu\rho ,\nu} + h_{\nu\rho ,\mu} - h_{\mu\nu , \rho}\right) - \frac{1}{d-2} h^{\lambda\rho} h_{\mu\nu}(\ln h)_{,\rho}$ where
$h = \det h^{\mu\nu}$.  If $d = 2$, then the equation of motion for $h^{\mu\nu}$ is consistent only if $h_{,\lambda} = 0$ and then if this
condition is satisfied, $G_{\mu\nu}^\lambda = \frac{1}{2} h^{\lambda\rho}\left(h_{\mu\rho ,\nu} + h_{\nu\rho ,\mu} - h_{\mu\nu ,\rho}\right)+h_{\mu\nu}X^{\lambda}$
where $X^{\lambda}$ is arbitrary.}
The case $d = 2$ is discussed in ref. [21, 22] where it is shown that in $2d$ the equation of motion does
not fix $\Gamma_{\mu\nu}^\lambda$ in terms of $g_{\mu\nu}$ unambiguously.  In the next section we will define variables in terms of the independent
fields $h^{\mu\nu}$ and $G_{\mu\nu}^\lambda$ that will facilitate the canonical analysis of $S_d$.  In the following section
this analysis will be performed using the Dirac constraint procedure [23-28].  The gauge invariance of the action
is then discussed.

The principle difference between this treatment and that of refs. [10, 11] where the first order form of the EH action has also
been considered is that not only primary and secondary, but also tertiary constraints arise.  This is because we do not use
constraint equations arising from equations of motion not involving time derivates to eliminate fields when these equations of motion
correspond to first class constraints.  Such tertiary constraints should be expected to occur, as only if tertiary first class constraints are
present can a gauge transformation involving the second derivative of the gauge function arise when the formalism of refs. [29, 30, 31]
is used to derive the gauge invariance present in a theory.  The diffeomorphism invariance of $S_d$ in eq. (1) is
\newpage
$$ \delta\,G_{\mu\nu}^\lambda = - \partial_{\mu\nu}^2 \theta^\lambda + \frac{1}{2} \left( \delta_\mu^\lambda \partial_\nu + \delta_\nu^\lambda
\partial_\mu\right)\partial_\rho\theta^\rho - \theta^\rho\partial_\rho G_{\mu\nu}^\lambda + G_{\mu\nu}^\rho \partial_\rho\theta^\lambda$$
$$-\left(G_{\mu\rho}^\lambda \partial_\nu + G_{\nu\rho}^\lambda
\partial_\mu\right)\theta^\rho \eqno(5a)$$
$$\delta h^{\mu\nu} = h^{\mu\lambda} \partial_\lambda \theta^\nu + h^{\nu\lambda} \partial_\lambda \theta^\mu -
\partial_\lambda \left(h^{\mu\nu} \theta^\lambda\right) \eqno(5b)$$
(which follow from
$$\begin{array}{l}\delta g^{\mu\nu}  =  -\theta^\lambda \partial_\lambda g^{\mu\nu} + g^{\mu\lambda} \partial_\lambda \theta^\nu +
g^{\nu\lambda} \partial_\lambda \theta^\mu \nonumber \\
 \delta\Gamma_{\mu\nu}^\lambda  =  -\partial_\mu \partial_\nu \theta^\lambda  + \Gamma_{\mu\nu}^\rho \partial_\rho \theta^\lambda
- \theta^\rho\partial_\rho \Gamma^\lambda_{\mu\nu} - \left(\Gamma_{\mu\rho}^\lambda \partial_\nu + \Gamma_{\nu\rho}^\lambda \partial_\mu\right)
\theta^\rho \nonumber \\
 \quad\delta g  =  g\,g^{\alpha\beta}\delta g_{\alpha\beta}\,.\,)\nonumber\end{array}$$
The second derivative of the ``gauge parameter'' $\theta^\lambda$ appears in eq.(5a); if there were no
tertiary constraints in the theory, such second derivatives would not be generated by the first class constraints. We will pay special attention to how
first class constraints generate gauge invariances, when using the approach of Henneaux, Teitelboim and Zanelli [30, 31, 25] and of Castellani [29].

We now rewrite $S_d$ in terms of variables that simplify the canonical analysis.

\section{Choosing Variables}

After an integration by parts (dropping the surface term) and defining $h = h^{00}$, $h^i = h^{0i}$, $\pi = -G_{00}^0$,
$\pi_i = -2G_{0i}^0$, $\pi_{ij} = -G_{ij}^0$, $\xi^i = -G_{00}^i$, $\xi^i_j = -2G_{j0}^i$, $\xi_{jk}^i = -G_{jk}^i$, we find that
$$\hspace{-10mm}{\cal{L}}_d = \pi h_{,0} + \pi_i h_{,0}^i +\pi_{ij} h_{,0}^{ij} + A + \xi^i A_i + \xi_j^i A_i^j + \xi_{jk}^i A_i^{jk}$$
$$\,\,\,\,\,\,+ \frac{1}{4} \left(\frac{1}{d-1} \xi_k^k \xi_\ell^\ell - \xi_\ell^k \xi_k^\ell\right) h + \left(\frac{1}{d-1} \xi_{ki}^k
\xi_{\ell j}^\ell - \xi_{\ell j}^k \xi_{kj}^\ell\right)h^{ij}$$
$$+\left(\frac{1}{d-1} \xi_k^k \xi_{\ell i}^\ell - \xi_\ell^k \xi_{ki}^\ell\right)h^i\eqno(6)$$
where
$$\hspace{-47mm} {A = \frac{2-d}{d-1} \left[ h\pi^2 + h^i \pi\,\pi_i + \frac{1}{4}  h^{ij} \pi_i \pi_j\right]}\eqno(7)$$
$$ \hspace{-7.5cm}A_i = h_{,i} - h\pi_i - 2h^j \pi_{ij}\eqno(8)$$
$$\hspace{-14mm} A_i^j = h_{,i}^j + \frac{1}{d-1} h\pi\,\delta_i^j + \frac{1}{2(d-1)} h^k\pi_k \delta_i^j - \frac{1}{2} h^j\pi_i - h^{jk} \pi_{ik}\eqno(9)$$
$$ A_i^{jk} = h_{,i}^{jk} + \frac{1}{d-1} \pi\left(\delta_i^j h^k + \delta_i^k h^j\right) + \frac{1}{2(d-1)}
\left(\delta_i^j h^{k\ell} + \delta_i^k h^{j\ell}\right)\pi_\ell\;\;.\eqno(10)$$

Eq. (6) can be further simplified by first decomposing $\xi_j^i$ into its trace $t$ and its traceless part $\overline{\xi}_j^i$,
$$\xi_j^i = \overline{\xi}_j^{\,i} + \frac{1}{d-1} \,\delta_j^i t\eqno(11)$$
followed by a shift to separate $\overline{\xi}_j^i$ from $\xi_{jk}^i$
$$\overline{\xi}_\ell^k = \overline{\zeta}_\ell^k - \frac{2}{h} \left(\xi_{\ell m}^k - \frac{1}{d-1} \,\delta_\ell^k \xi_{jm}^j\right)h^m\eqno(12)$$
so that
$$\hspace{-1cm}{\cal{L}}_d = \pi h_{,0} + \pi_i h_{,0}^i + \pi_{ij} h_{,0}^{ij} + A + \xi^i A_i + \overline{\zeta}_i^{\,j} A_j^i +
\frac{t}{d-1} A_i^i\eqno(13)$$
$$ + \xi_{jk}^i \left(A_i^{jk} - \frac{1}{h} \left(h^j A_i^k + h^kA_i^j\right) + \frac{1}{(d-1)h} \left(h^j \delta_i^k + h^k
\delta_i^j\right)A_\rho^\rho\right)$$
$$- \frac{1}{4}\, \overline{\zeta}_j^{\,i}\, \overline{\zeta}_i^{\,j} h + \left(\frac{1}{d-1} \xi_{ki}^k \xi_{\ell j}^\ell - \xi_{\ell i}^k \xi_{kj}^\ell \right) \left(h^{ij} - h^i h^j/h\right).$$

In eq. (13) we see that the effect of making the change of variables defined in eqs. (11, 12) is two fold.  First of all, $t$ appears only linearly in eq. (13) which
means that eventually its presence will lead to a secondary constraint ($\chi$, defined in eq. (16) below) that turns out to be first class.  Secondly, the traceless quantity
$\overline{\zeta}^i_j$ and the quantity $\xi_{jk}^i$ are ``decoupled'' in eq. (13) as eq. (12) is effectively a ``completing the square'' operation.  These two quantities
only enter eq. (13) quadraticly, resulting eventually in two second generation constraints (see eqs. (23, 24) below) which are necessarily second class.  The tracelessness of
$\overline{\zeta}^i_j$ will require special consideration when defining its Poisson Bracket with its conjugate momentum (see. eq. (18) below).  Indeed, one might supplement
${\cal{L}}_d$ in eq. (13) with a terms $\lambda\overline{\zeta}^i_i$ where $\lambda$ is a Lagrange multiplier used to generate a constraint that ensures that
$\overline{\zeta}_j^i$ is traceless, but this is not necessary.

We now perform the change of variables
$$H^{ij}  =  h\,h^{ij} - h^i h^j \,\,\,\left(H_{ik} H^{kj} \equiv \delta_i^j\right)\eqno(14a)$$
$$\hspace{-3.98cm}\pi_{ij}  =  h \Pi_{ij}\eqno(14b)$$
$$\hspace{-2.65cm}\pi_i  =  \Pi_i - 2\Pi_{ij} h^j\eqno(14c)$$
$$\hspace{-.83cm}\pi  =  \Pi + \frac{\Pi_{ij}}{h} \left(H^{ij} + h^i h^j\right);\eqno(14d)$$
$$\hspace{-1.3cm}\overline{t}  =  \frac{1}{d-1} \left(t + \frac{2}{h} \xi_{ij}^i h^j\right)\eqno(14e)$$
$$\hspace{-2.5cm}\overline{\xi}^{\,i}  =  \xi^i - \xi_{jk}^i h^{jk}/h\eqno(14f)$$
so that
$$\hspace{-6cm}{\cal{L}}_d = \Pi h_{,0} + \Pi_i h_{,0}^i+ \Pi_{ij} H_{,0}^{ij} + \overline{\xi}^{\,i} \chi_i + \overline{t} \chi\eqno(15)$$
$$\hspace{-3cm}+ \frac{2-d}{d-1} \left[ h\Pi^2 + h^i \Pi\,\Pi_i + \frac{1}{4h} \left(H^{ij} + h^ih^j\right)\Pi_i\Pi_j\right.$$
$$\hspace{-4cm}+ \frac{1}{h} \left( H^{ij}H^{k\ell} + H^{ik}h^jh^\ell\right)\Pi_{ij}\Pi_{k\ell}$$
$$\hspace{-4cm}\left. + \frac{1}{h} \left(h^i H^{k\ell} - H^{ik}h^\ell\right) \Pi_i\Pi_{k\ell} + 2H^{ij}\Pi\,\Pi_{ij}\right]$$
$$\hspace{-5cm}+ \overline{\zeta}_j^{\,i} \left( h_{,i}^j - \frac{1}{2} h^j \Pi_i - H^{jk} \Pi_{ik}\right) - \frac{h}{4} \,\,\overline{\zeta}_j^{\,i}
\,\,\overline{\zeta}_i^{\,j}$$
$$+ \xi_{jk}^i \left[ \frac{1}{h} H_{,i}^{jk} - \frac{1}{h} H^{jk} \Pi_i + \frac{1}{2(d-1)h} \left(\delta_i^j H^{k\ell} +
\delta_i^k H^{j\ell}\right)\left(\Pi_\ell - 2h^m \Pi_{\ell m}\right)\right.\nonumber$$
$$\left. \left.+ \frac{1}{h} \left( h^j H^{kp} + h^k H^{jp}\right)\Pi_{ip}\right]	+ \frac{1}{h} H^{ij}
\left(\frac{1}{d-1} \xi_{ki}^k \xi_{\ell j}^\ell - \xi_{\ell i}^k \xi_{kj}^\ell\right)\right]$$
where
$$\chi_i = h_{,i} - h\Pi_i \eqno(16)$$
$$\chi = h_{,i}^i + h\Pi\,\,.\eqno(17)$$
With the choice of variables used to express ${\cal {L}}_d$ in the form of eq. (15), we can now examine its canonical structure.

\section{The Canonical Structure}

We begin by noting that the momenta conjugate to $h$, $h^i$ and $H^{ij}$ are given by $\Pi$, $\Pi_i$ and $\Pi_{ij}$ respectively,
while the momenta conjugate to $\overline{t}$, $\overline{\xi}^{\,i}$, $\overline{\zeta}_j^i$ and $\xi_{jk}^i$ (which we denote by
$I\!\!P$, $I\!\!P_i$, $\overline{I\!\!P}_i^{\,j}$ and $I\!\!P_i^{\,jk}$ respectively) all vanish.  The fundamental Poisson Brackets (PBs)
are standard except for
$$\left\lbrace \overline{\zeta}_j^{\,i} (\vec{r}, t), \overline{I\!\!P}_\ell^k (\vec{r}^{\;\prime},t)\right\rbrace =
\left(\delta_\ell^i \delta_j^k - \frac{1}{d-1} \delta_j^i \delta_\ell^k\right)\delta^3 (\vec{r} - \vec{r}^{\;\prime})\eqno(18)$$
where the second term on the right hand side of eq. (18) ensures consistency with the tracelessness of $\overline{\xi}_j^{\,i}$ and
$\overline{I\!\!P}_\ell^{\,k}$. (As mentioned above, one could use a Lagrange multiplier in the action to generate the constraint $\overline{\zeta}_i^{\,i} = 0$ in
which case eq. (18) actually becomes a Dirac Bracket (DB) associated with this constraint and the gauge condition $\overline{I\!\!P}^i_i = 0$ {as discussed below.)

It is now possible to read off the canonical Hamiltonian ${\cal{H}}_c$ from eq. (15) as ${\cal{L}}_d$ is of the form
$${\cal{L}}_d = \Pi h_{,0} + \Pi_i h_{,0}^i + \Pi_{ij} H_{,0}^{ij} - {\cal{H}}_c\;\;.\eqno(19)$$

The Dirac constraint formalism [23-28] now should be implemented as there are the obvious primary constraints
$$I\!\!P = I\!\!P_i = \overline{I\!\!P}_i^{\,j} = I\!\!P_i^{jk} = 0 \;\;.\eqno(20a-d)$$

The primary constraints $I\!\!P = I\!\!P_i = 0$ immediately give rise to the secondary constraints
$$\chi = \chi_i = 0\;\;.\eqno(21a,b)$$
Test functions $f$ and $g$ can be used to compute [32] the PB of $\chi_i$ and $\chi$;
$$\hspace{-3cm}\int d\vec{r} d\vec{r}^{\;\prime} f(\vec{r}) \left\lbrace \chi_i (\vec{r}, t), \chi(\vec{r}^{\;\prime},t)\right\rbrace
g(\vec{r}^{\;\prime}) \equiv f\left\lbrace \chi_i, \chi\right\rbrace g\nonumber$$
$$ = f\left\lbrace h_{,i,} \Pi\right\rbrace hg - f \Pi_i \left\lbrace h_,\Pi\right\rbrace hg - fh\left\lbrace\Pi_i, h_{,j}^j\right\rbrace g\nonumber$$
$$ \hspace{-3.5cm}= - f_{.i} hg - f h\Pi _i g + fh(-g_{,i})\nonumber$$
$$\hspace{-1cm} =\int d\vec{r} f(\vec{r}) \left(h_{,i}(\vec{r},t) - h(\vec{r}, t)\Pi_i(\vec{r}, t)\right)g(\vec{r})\nonumber$$
$$\hspace{-7cm}= f\chi_i g\nonumber$$
or, more compactly,
$$\left\lbrace \chi_i, \chi \right\rbrace = \chi_i\;\;. \eqno(22)$$
It is also apparent that $\left\lbrace \chi_i, \chi_j\right\rbrace = 0 = \left\lbrace \chi , \chi \right\rbrace$.  Consequently
$I\!\!P$, $I\!\!P_i$, $\chi$ and $\chi_i$ are all candidates for being first class constraints but no conclusion about the class of $\chi$ and
$\chi_i$ can be drawn until the complete set of constraints is determined as it is possible that $\chi$ and/or $\chi_i$ do not have a weakly vanishing
PB with a tertiary constraint.

In ref. [10,11] equations of motion that do not involve time derivatives were used to eliminate fields from the initial action. Two of these equations
are the constraints $\chi$ and $\chi_i$ (ie, the trace of eq. (A3) and eq. (A4) of ref. [11]).  Consequently, tertiary constraints cannot arise in the
approach of refs. [10,11].

The constraints $\overline{I\!\!P}_j^{\,i} = I\!\!P_i^{jk} = 0$ also lead to secondary constraints $\Theta_j^i$ and $\Theta_i^{jk}$ respectively.  These are linear
in $\overline{\zeta}_j^{\,i}$ and $\xi_{jk}^i$ and so all these constraints are immediately seen to be second class.  These
secondary constraints correspond to eq. (A2) and the traceless part of eq. (A.3) in ref. [11].  As they are second class,
they can be used to eliminate fields from the action provided PBs are replaced by the appropriate DB.

If $d = 2$, $\Theta_j^i$ does not arise and $\Theta_i^{jk}$ reverts to being a single first class constraint, making the canonical
analysis [12-17] considerably simpler than when $d > 2$.

Once the DB replaces the PB, $\Theta_j^i$ and $\Theta_i^{jk}$ are effectively eliminated from the theory and hence one need not worry
about the fact that the constraints $\chi$, $\chi_i$ have non-vanishing PB with $\Theta_j^i$ and $\Theta_i^{jk}$ [33].

The portions of the Hamiltonian ${\cal{H}}_c$ that contributes to $\Theta_j^i$ and $\Theta_i^{jk}$ are of the form
$$\hspace{-3.9cm}A = -\overline{\zeta}_j^{\,i} \lambda_i^j + \frac{h}{4} \overline{\zeta}_j^{\,i} \overline{\zeta}_i^{\,j}\eqno(23)$$
$$B = -\xi_{jk}^i \sigma_i^{jk} - \frac{H^{ij}}{h} \left(\frac{1}{d-1} \xi_{ki}^k \xi_{\ell j}^\ell -
\xi_{\ell i}^k \xi_{kj}^\ell\right)\eqno(24)$$
$$\hspace{-1.9cm}\equiv - \xi_{jk}^i \sigma_i^{jk} - \xi_{\ell m}^k M_{k\quad c}^{\ell m\;\,de} \xi_{de}^c\;\;\nonumber$$
where
$$\hspace{-4.2cm} \lambda_i^j = h_{,i}^j - \frac{1}{2} h^j \Pi_{i} -  H^{jk} \Pi_{ik}\nonumber$$
$$
\sigma_i^{jk} =  \frac{1}{h}  H^{jk}_{,i} -  \frac{1}{h} H^{jk} \Pi_i + \frac{1}{2(d-1)h} \left(\delta_i^j H^{k\ell} +
\delta_i^k H^{j\ell}\right)\nonumber$$
$$\left(\Pi_\ell - 2h^m \Pi_{\ell m}\right)+ \frac{1}{h} \left( h^j H^{kp} + h^k H^{jp}\right)\Pi_{ip}\nonumber$$
The constraints $\Theta_j^i$ and $\Theta_i^{jk}$ that arise from $A$ and $B$ can be solved, yielding
$$\overline{\zeta}_j^{\,i} = \frac{2}{h} \left(\lambda_j^i - \frac{1}{d-1} \delta_j^i \lambda_k^k\right)\eqno(25)$$
$$\xi_{jk}^i = - \frac{1}{2}\left(M^{-1}\right)_{jk\;\;mn}^{i\quad \ell}  \sigma_\ell^{mn}\eqno(26)$$
where
$$\hspace{-.9cm}\left(M^{-1}\right)_{yz\;\;\ell m}^{x\quad k}
 = - \frac{h}{2}
\left[ \left( H_{\ell y} \delta_z^k \delta_m^x +  H_{\ell z} \delta_y^k \delta_m^x +
 H_{m y} \delta_z^k \delta_\ell^x +  H_{mz} \delta_y^k \delta_\ell^x\right)\right.\nonumber$$
$$\qquad\left. + \frac{2}{d-2} \left(H^{kx} H_{\ell m} H_{yz}\right) - H^{kx}\left(H_{\ell z}H_{my} + H_{\ell y}H_{mz}\right)\right].\eqno(27)$$
(Decomposing $\xi_{jk}^i$ so that $\xi_{ij}^i = t_j$ and $H^{ij}\xi_{ij}^k = s^k$ by setting
$$\xi_{jk}^i = \overline{\eta}_{jk}^i + \frac{1}{(d-2)(d+1)} \left[ dH_{jk}s^i - \left(\delta_j^i H_{k\ell} +
\delta_k^i H_{j\ell}\right)s^\ell\right.\nonumber$$
$$\left. -2H^{i\ell}H_{jk}t_\ell + (d-1)\left(\delta_j^i t_k + \delta_k^i t_j\right)\right]\nonumber$$
does not really simplify the canonical analysis.)

One could now substitute eqs. (25, 26) into ${\cal{H}}_c$ provided the appropriate DBs are subsequently used.  To illustrate how these DBs are
worked out, we note that $A$ and $B$ of eqs. (23, 24) are both of the form (with $I = 1,2$)
$$/\!\!\!C = f_I^a \left(q_i, p_i\right) Q_I^a - \frac{1}{2} Q_I^a g_{IJ}^{ab} (q_i)Q_J^b\eqno(28)$$
where $q_i$ and $p_i$ are canonically conjugate variables.  The momentum $I\!\!P_I^a$ conjugate to $Q_I^a$ vanishes and this primary
constraint leads to the secondary constraint
$$\theta_I^a = f_I^a\left(q_i, p_i\right) - g_{IJ}^{ab}(q_i)Q_J^b\eqno(29)$$
with $\Theta_I^a = \left(I\!\!P_I^a, \theta_I^a\right)$ being second class, as
$$\left\lbrace I\!\!P_I^a, \theta_J^b\right\rbrace = g_{IJ}^{ab} \delta_{IJ}\eqno(30)$$
$$\left\lbrace \theta_I^a, \theta_J^b \right\rbrace \equiv M_{IJ}^{ab}\;\;.\eqno(31)$$
In order to form the DB
$$\left\lbrace A, B \right\rbrace^* = \left\lbrace A, B \right\rbrace -
\left\lbrace A,\Theta_I^a \right\rbrace (d^{-1})_{I\,\,J}^{a\,\,b}
\left\lbrace \Theta_J^b, B \right\rbrace \eqno(32)$$
one needs the inverse of
$$ d_{I\,\,J}^{a\,\,b} =
\left( \begin{array}{cccc}
0 & g_1 & 0 & 0\\
-g_1 & M_{11} & 0 &M_{12}\\
0 & 0 & 0 & g_2\\
0 & -M_{12} & -g_2 & M_{22}\end{array} \right)\,\,.\eqno(33)$$
Using the standard relation
$$\hspace{-4cm}\left(\begin{array}{cc}
A & B\\
C & D\end{array}\right)^{-1}  =
\left[\left(\begin{array}{cc}
I & B\\
0 & D\end{array}\right)\,\,
\left(\begin{array}{cc}
A-BD^{-1}C & 0\\
D^{-1}C & I\end{array}\right)\right]^{-1}\eqno(34)$$
$$\hspace{2.6cm} = \left( \begin{array}{cc}
(A-BD^{-1}C)^{-1} & -(A-BD^{-1}C)^{-1}BD^{-1}\\
-D^{-1}C(A-BD^{-1}C)^{-1} & D^{-1}C(A-BD^{-1}C)^{-1}BD^{-1} + D^{-1}
\end{array}\right)\nonumber $$
we see that
$$d^{-1} = \left(\begin{array}{cccc}
g_1^{-1} M_{11} g_1^{-1} & -g_1^{-1} & g_1^{-1} M_{12} g_2^{-1}&0\\
g_1^{-1} & 0 & 0 & 0\\
-g_2^{-1} M_{12} g_1^{-1} & 0 & g_2^{-1} M_{22} g_2^{-1} & -g_2^{-1}\\
0 & 0 & g_2^{-1} & 0\end{array}\right)\;\;.\eqno(35)$$
(In fact this inverse is not unique, as is discussed on pg. (66) of ref. [27], but the form given in eq. (35) suits our purpose.)
From eqs.(32, 35) we find that the non vanishing DBs are
$$\hspace{-2.2cm}\left\lbrace q_i, p_j\right\rbrace^* = \delta_{ij}\eqno(36)$$
$$\left\lbrace q_i, Q_{Ia}\right\rbrace^* = \left\lbrace q_i, \theta_I^c\right\rbrace (g^{-1})_{IJ}^{ca} \delta_{IJ}\eqno(37)$$
$$\left\lbrace p_i, Q_{Ia}\right\rbrace^* = \left\lbrace p_i, \theta_I^c\right\rbrace (g^{-1})_{IJ}^{ca} \delta_{IJ}\eqno(38)$$
$$\hspace{.5cm}\left\lbrace Q_I^a, Q_J^b\right\rbrace^* = \left(g^{-1}\right)_{IK}^{am} M_{KL}^{mn} (g^{-1})_{LJ}^{nb}\eqno(39)$$
$$\left\lbrace Q_I^a, I\!\!P_J^b\right\rbrace^* = 0\;\;.\eqno(40)$$
If we use these DBs, it is now possible to employ the second class constraint $\theta_I^a = 0$ to write $/\!\!\!C$ in eq. (28) as
$$ /\!\!\!C = \frac{1}{2} Q_I^a g_{I\,\,J}^{a\,\,b} (q_i)Q_J^b\;\;\eqno(41)$$
and so by eqs. (36-40) for any function $Z(q_i, p_i)$
$$\left\lbrace Z\left(q_i , p_i\right), /\!\!\!C  \right\rbrace^*  = Q_I^a \left\lbrace Z\left(q_i , p_i\right), f_I^a\left(q_i , p_i\right)\right\rbrace\eqno(42)$$
$$\hspace{3cm}- \frac{1}{2} Q_I^a \left\lbrace Z\left(q_i , p_i\right), g_{IJ}^{ab}(q_i)\right\rbrace Q_j^b\;\;.\nonumber$$

Using eq.(42) it is quite easy to show that (with $H_c = \int d^{d-1}x{\cal{H}}_c$)
$$\left\lbrace \chi , H_c\right\rbrace^* = {\cal{H}}_c - 2 \overline{\xi}^{\,i}\chi_i - \overline{t}\chi + \delta^i_{,i}\eqno(43)$$
where
$$\delta^i = \frac{d-2}{d-1} \left[\frac{\Pi_{k\ell}}{h} \left(h^iH^{k\ell} - H^{ik}h^\ell\right) + \Pi\,h^i + \frac{1}{2h}
\left(H^{ij} +  h^ih^j\right)\Pi_j\right]\nonumber$$
$$+ \frac{1}{2}\, \overline{\zeta}^{\,i}_j h^j + \frac{1}{h} \xi_{jk}^i H^{jk} - \frac{1}{d-1}\, \frac{H^{ij}}{h} \, \xi_{jk}^k \eqno(44)$$
so that a tertiary constraint arises that differs from the
Hamiltonian ${\cal{H}}_c$ by a linear combination of secondary constraints and a total derivative.

In fact, the Hamiltonian that follows from eq. (15) can be written as
$${\cal{H}}_c = \frac{1}{h}\left(\tau + h^i\tau_i\right) + \frac{3}{2(d-2)} \frac{1}{h^2} H_{k\ell}H_{,i}^{k\ell} H^{ij} \chi_j - \frac{3d-5}{4(d-2)}
\frac{1}{h^3} \,H^{ij} \chi_i\chi_j\nonumber$$
$$\hspace{-.6cm}-\frac{3}{h^2}H^{ij}\chi_i\Pi_j - \left(\frac{1}{h^2} H^{ij}\chi_i\right)_{,j} + \frac{2}{h^2}H_{\;\;,i}^{ij} \chi_j + \frac{h^i}{h}\,\chi_{,i}\eqno(45)$$
$$\hspace{-.6cm}-\frac{1}{h^2} h^i\,h_{,i}^j\chi_j - 2 \left(\frac{d-2}{d-1}\right) \,\frac{1}{h} H^{mn}\Pi_{mn}\chi - \frac{1}{h} h^i\Pi\chi_i\nonumber$$
$$\hspace{-.7cm}- \frac{H^{mn}\Pi_{mn}}{h^2} \,h^i\chi_i - \frac{2}{h^2} h^kH^{ij} \Pi_{ik}\chi_j + \frac{d-3}{d-1} \Pi\chi\nonumber$$
$$\hspace{-.7cm}+ \frac{1}{d-1}\,\frac{\chi^2}{h} - \frac{1}{d-1} \,\frac{1}{h} h^i \Pi_i\chi - \overline{\xi}^{\,i}\chi_i - \overline{t}\chi\nonumber$$
where
$$\hspace{-3cm}\tau = H_{\,\,,ij}^{ij} - \frac{1}{2} H_{\,\,,n}^{mi} H_{ij} H_{\,\,,m}^{nj} -\frac{1}{4} H^{ij} H_{mn,i} H_{\,\,,j}^{mn} \eqno(46)$$
$$\hspace{1cm}- \frac{1}{4(d-2)}H^{ij}H_{k\ell} H_{,i}^{k\ell} H_{mn} H_{,j}^{mm} + H^{ij} H^{k\ell}\left(\Pi_{ij}\Pi_{k\ell} - \Pi_{ik}\Pi_{j\ell}\right)\nonumber$$
and
$$\tau_i = 2\left(H^{mn}\Pi_{mi}\right)_{,n} - H^{mn}\Pi_{mn,i} - \left(H^{mn}\Pi_{mn}\right)_{,i}\; .\eqno(47)$$

It is evident from the form of the Hamiltonian given in eq. (45) that the secondary constraints $\chi$ and $\chi_i$ imply tertiary constraints
$\tau$ and $\tau_i$.  We immediately see that $\chi$ and $\chi_i$ have vanishing PBs with $\tau$ and $\tau_i$.  Furthermore, we find that the PBs amongst
$\tau$ and $\tau_i$ are [61]
$$\left\lbrace \tau_i (\vec{x}), \tau_j(\vec{y})\right\rbrace = - \partial_j^x \delta\left(\vec{x} - \vec{y}\right) \tau_i
(\vec{y}) + \tau_j(\vec{x})
\partial_i^y\delta\left(\vec{x} - \vec{y}\right)\eqno(48)$$
$$\hspace{-2cm}\left\lbrace \tau (\vec{x}), \tau(\vec{y})\right\rbrace =  \partial_i^x \delta\left(\vec{x} - \vec{y}\right) H^{ij}(\vec{y})\tau_j(\vec{y})\eqno(49)$$
$$-H^{ij}(\vec{x}) \tau_j(\vec{x})\partial_i^y \delta\left(\vec{x} - \vec{y}\right)\nonumber$$
$$\left\lbrace \tau_i (\vec{x}), \tau(\vec{y})\right\rbrace = - \partial_i^x \delta\left(\vec{x} - \vec{y}\right) \tau(\vec{y}) + \tau(\vec{x})
\partial_i^y\delta\left(\vec{x} - \vec{y}\right)\,.\eqno(50)$$
(Using test functions as in the derivation of eq. (22) is useful in demonstrating eqs. (48-50).)  The PB algebra of eqs. (48-50) is that of ref. [11] even
though in this reference the constraints are distinct from those of eqs. (46, 47).  Furthermore the ``ADM constraints'' appearing in ref. [11] are secondary constraints
derived from an ``ADM action'' found by substitution of solutions to the true secondary constraints, both first class ($\chi$ and $\chi_i$) and second class
($\Theta_j^i$ and $\Theta_i^{jk}$), into the first order EH action of eq. (2).  After this substitution, the EH action becomes the ``ADM action''.  In contrast, the constraints
$\tau$ and $\tau_i$ appearing in eqs. (48-50) are true first class tertiary constraints that follow from the secondary first class constraints
$\chi$ and $\chi_i$ and the EH action.  It is evident that because of the form of ${\cal{H}}_c$ in eq. (45) no further constraints of a generation beyond the third arise.

We thus have the complete constraint structure of the EH action of eq. (2).  Initially there are $d(d+1)^2$ variables in phase space ($h^{\mu\nu}$,
$G_{\mu\nu}^\lambda$ and their conjugate momenta).  There are $d(d+1)$ primary second class constraints identified with the canonical momenta conjugate to $h$,
$h^i$ and $H^{ij}$.  In addition there are $d(d^2 - 3)/2$ primary second class constraints associated with the vanishing of the momenta conjugate
to $\overline{\zeta}^{\,i}_j$ and $\xi^i_{jk}$ which in turn lead to the $d(d^2 - 3)/2$ secondary second class constraints $\Theta_j^i$ and $\Theta_i^{jk}$.
Finally there are $3d$ first class constraints spanning three generations ($I\!\!P$, $I\!\!P_i$; $\chi$, $\chi_i$; $\tau$, $\tau_i$) which require $3d$ gauge
conditions.  In total then there are $\displaystyle{d(d + 1) + \frac{d(d^2-3)}{2} + \frac{d(d^2-3)}{2} + 3d + 3d = d(d^2 + d + 4)}$ restrictions on the system leaving
$d(d + 1)^2 - d(d^2 + d + 4) = d(d - 3)$ independent degrees of freedom in phase space.  In $d = 3$ this is zero; if $d = 4$ this is four which corresponds
to the two polarizations of the graviton and their conjugate momenta.

In the ADM approach to the first order action of eq. (2) [10, 11] just $d = 4$ dimensions is considered.  Only six of the ten components of the metric ($g_{ij}$) are taken to be dynamical,
the remaining four form the non-dynamical ``lapse'' ($N$) and ``shift'' ($N_i$) functions.  All thirty constraint equations $\chi = \chi_i = \Theta_j^i = \Theta_i^{jk} = 0$ are used to
eliminate fields in the EH action.  Once this is done the four fields $\Gamma_{00}^\mu$ disappear from this reduced ``ADM action'' and are not considered to be
dynamical.  There are then six components of $\Gamma_{\mu\nu}^\lambda$ remaining which are used to form the momenta conjugate to the six dynamical components of the
metric.  The four ADM constraints derived from the ADM action when combined with their associated gauge conditions leave just the two degrees of freedom present in the
metric plus their conjugate momenta.  We thus see how the analysis of this paper, which exclusively uses the Dirac constraint formalism, is related to the more conventional
ADM approach to the first order action of eq. (2).  It is apparent that the ADM approach initially resembles that of ref. [34] in that equations of motion not containing
time derivatives are used to eliminate fields; only after this has been done in Dirac's constraint formalism invoked.  This approach cannot
lead to tertiary constraints, which are necessary if one is to obtain the term $-\partial^2_{\mu\nu}\theta^\lambda$ in
the gauge transformation of eq. (5a) from a generator constructed from the first class constraints in
the theory.  The relationship between the Dirac constraint formalism and that of [34] is further discussed in ref. [35].

\section{The Gauge Transformation}

There are several ways of deriving the form of the gauge transformation that leaves the action invariant from the first class constraints present in the theory [29-31].
These methods have been applied to the complete constraint analysis of the second order EH action to show that its gauge symmetry is in fact diffeomorphism symmetry [8, 9],
while the second order ADM action is invariant only under a diffeomorphism if there is a field dependent gauge function [29,36,37] (which would alter the group properties of
the gauge symmetry [37]).

We first will examine the symmetries of the first order EH action from the Lagrangian point of view, then give an (incomplete) discussion based on the approachs of
Henneaux, Teitelboim and Zanelli (HTZ) appearing in [30, 31, 25] and of Castellani [29].

In discussing the invariances of the action from the Lagrangian point of view, we will adapt the approach of [38] in a way that retains manifest covariance.  In general, if an
action depends on a field $\phi_A(x)$, then variation of the action under a variation $\delta \phi^A$ of the field is given by
$$\delta S = \int \,d^d x \left[- \partial_\mu \,\frac{\delta{\cal{L}}}{\delta\left(\partial_\mu\phi_A\right)}  + \frac{\delta\cal{L}}{\delta\phi_A} \right]
\delta\phi_A \equiv \int d^d x L^A \delta\phi^A\eqno(51)$$
provided $\delta\phi^A$ vanishes at infinity.  ($L^A$ is the ``Euler-Lagrange'' (EL) derivative.)  We now consider variations of the form
$$\delta\phi^A = \sum_{s=0}^N \left[ (-1)^s \frac{\partial^s}{\partial x_{\mu_{1}}\ldots \partial x_{\mu_{s}}} \,\eta^B (x)\right] \rho_{\mu_1 \ldots \mu_s}^{AB}\,(x)\eqno(52)$$
where $\eta^B$ are unspecified ``gauge functions'' and the quantities $\rho_{\mu_{1} \ldots \mu_{s}}^{AB}$ are to be determined.  Substitution of eq. (52) into eq. (51) yields
$$\delta S = \int \,d^d x \,\eta^B \sum_{s=0}^N \left[ \frac{\partial^s}{\partial x_{\mu_{1}}\ldots \partial x_{\mu_{s}}} \, \rho_{\mu{_1} \ldots \mu{_s}}^{AB}\,L^A\right]\eqno(53)$$
which vanishes even if $\phi^A$ does not satisfy the equations of motion provided
$$\sum_{s=0}^N \left[ \frac{\partial^s}{\partial x_{\mu_{1}}\ldots \partial x_{\mu_{s}}} \, \rho_{\mu_1 \ldots \mu_s}^{AB}\,L^A\right] = 0\eqno(54)$$
and surface terms are neglected.
This equation is used to determine $\rho_{\mu_1 \ldots \mu_s}^{AB}$ systematicly as illustrated below.  If an invariance of the action were known (so that we have
the explicit form of the functions $\rho_{\mu_{1}\ldots \mu_{s}}^{AB}$) then eq. (54) would give the associated Noether currents of the model.

Varying the fields appearing in eq. (2) we find that
$$\delta S_d = \int d^d x\left[\left(G_{\mu\nu , \lambda}^\lambda + \frac{1}{d-1} G_{\lambda\mu}^\lambda G_{\sigma\nu}^\sigma - G_{\sigma\mu}^\lambda G_{\lambda\nu}^\sigma\right)
\delta h^{\mu\nu}\right.\eqno(55)$$
$$+ \left( - h _{,\lambda}^{\mu\nu} + \frac{1}{d-1}\left(h^{\mu\rho}\delta_\lambda^\nu + h^{\nu\rho}\delta_\lambda^\mu\right)G_{\sigma\rho}^\sigma \right.\nonumber$$
$$-\left. \left.(h^{\mu\rho} G_{\lambda\rho}^\nu + h^{\nu\rho} G_{\lambda\rho}^\mu)\right) \delta G_{\mu\nu}^\lambda\right].\nonumber$$
Eq. (55) will be used to systematicly determine the invariances of $S_d$.  If the
fields $\phi_A$ of eq. (51) are identified with ($h^{\mu\nu}, G^\lambda_{\mu\nu}$), we can construct $\delta\phi_A$ from eq. (55) using a step-by-step procedure which amounts to
expanding $\rho^{AB}_{\mu_1\ldots\mu_5}$ in eq. (52) in powers of $\phi_A$.  We assume that we can expand $\delta\phi_A = {\displaystyle{\sum_{k=1}^N}} \delta^{(k)} \phi_A$ where
$\delta^{(k)}\phi_A$ contains $k$ factors of $\phi_A$. $\delta S$ in eq. (8) is then expressed as a series in powers of $\phi_A$ with each term in this expansion being set equal
to zero in order to fix $\delta^{(k)}\phi_A$ in terms of $\delta^{(k-1)}\phi_A$.  If this procedure terminates with $\delta^{(N+1)}\phi_A = 0$ then we have an invariance of the theory.

For example, in Yang-Mills theory, we have
$$ S_{ym} = -\frac{1}{4} \int d^4x \left(\partial_\mu A_\nu^a - \partial_\nu A_\mu^a + \epsilon ^{abc}A_\mu^b A_\nu^c\right)^2\nonumber$$
so that
$$ \delta S_{ym} = \int d^4x \left[\left( \partial_\mu \delta^{ab} + \epsilon^{apb} A_\mu^p\right)\left(\partial_\mu A_\nu^b - \partial_\nu A_\mu^b + \epsilon^{bcd} A_\mu^c A_\nu^d\right)
\right](\delta A_\nu^a) .\nonumber$$
If now $\delta A_\nu^a = \displaystyle{\sum_{k=0}^N} \delta^{(k)} A_\nu^a$, then $\delta S_{ym} = 0$ at each order in $A_\mu^a$ if
$$\int d^4x \left[ \partial_\mu \left(\partial_\mu A_\nu^a - \partial_\nu A_\mu^a\right)\right](\delta^{(0)} A_\nu^a) = 0\nonumber$$
$$\int d^4x \left[\left(\partial_\mu (\partial_\mu A_\nu^a - \partial_\nu A_\mu^a)\right)(\delta^{(1)}A_\nu^a)\right.\nonumber$$
$$ \hspace{2cm}+ (\epsilon^{apb}A_\mu^p)\left(\partial_\mu A_\nu^b - \partial_\nu A_\mu^b\right) (\delta^{(0)}A_\nu^a)\nonumber$$
$$\hspace{2cm}\left. + \left(\partial_\mu\left(\epsilon^{abc}A_\mu^b A_\nu^c\right) \right)(\delta^{(0)}A_\nu^a)\right] = 0\nonumber$$
etc.
These equations are automaticly satisfied if $\delta^{(0)} A_\nu^a = \partial_\nu\theta^a$, $\delta^{(1)} A_\nu^a = \epsilon^{abc}A_\nu^b\theta^c$ and $\delta^{(2)} A_\nu^a = 0$; thus
$\delta A_\mu^a = \partial_\mu\theta^a + \epsilon^{abc}A_\mu^b\theta^c$ is an invariance of the Yang-Mills action.

There is another more direct way of constructing $\delta\phi_A$ by using eq. (54) that is used in ref.[45].

Applying this approach to eq. (55), we find that $\delta S_d = 0$ provided
$$\int d^d x \left[ G_{\mu\nu , \lambda}^\lambda \delta^{(0)} h^{\mu\nu} + \left(-h^{\mu\nu}_{,\lambda}\right)\delta^{(0)} G_{\mu\nu}^\lambda \right] = 0 \eqno(56a)$$
and
$$\int d^d x \left[ G_{\mu\nu , \lambda}^\lambda \delta^{(i+1)} h^{\mu\nu} + \left(-h^{\mu\nu}_{,\lambda}\right)
\delta^{(i+1)} G_{\mu\nu}^\lambda + \left(\frac{1}{d-1}  G_{\lambda\mu}^\lambda G_{\sigma\nu}^\sigma -  G_{\sigma\mu}^\lambda G_{\lambda\nu}^\sigma\right)\delta^{(i)} h^{\mu\nu}\right.\nonumber$$
$$\hspace{-2cm}+ \left( \frac{1}{d-1} \left(h^{\mu\rho}\delta_\lambda^\nu + h^{\nu\rho}\delta_\lambda^\mu\right)G_{\sigma\rho}^\sigma \right.\nonumber$$
$$-\left.\left. (h^{\mu\rho}G_{\lambda\rho}^\nu + h^{\nu\rho}G_{\lambda\rho}^\mu)\right)\delta^{(i)} G_{\mu\nu}^\lambda\right] = 0\eqno(56b)$$
$$(i = 0, 1 \ldots , \; N - 1).$$

If $d = 2$, then eq. (56a) can be satisfied in two ways, first with
$$\delta^{(0)}\,h^{\mu\nu} = 0 \qquad\delta^{(0)} G_{\mu\nu}^\lambda = -\partial_{\mu\nu}^2 \theta^\lambda + \frac{1}{2}\left(\delta_\mu^\lambda \theta_{,\rho\nu}^\rho +
\delta_\nu^\lambda \theta_{,\rho\mu}^\rho\right)\eqno(57a,b)$$
and second with
$$\overline{\delta}^{\,(0)} h^{\mu\nu} = 0\qquad\overline{\delta}^{\,(0)} G_{\mu\nu}^\lambda = -\epsilon^{\lambda\tau}\xi_{\mu\nu ,\tau}\;.\eqno(58a,b)$$
(In eq. (58), $\xi_{\mu\nu}$ is a symmetric tensor and $\epsilon^{01} = - \epsilon^{10} = 1$.) Eq. (57) can be used in conjunction with eq. (56b) to yield
$$\hspace{-2.2cm}\delta^{(1)} h^{\mu\nu} = h^{\mu\lambda} \partial_\lambda\theta^\nu + h^{\nu\lambda}\partial_\lambda\theta^\mu - \partial_\lambda\left(h^{\mu\nu}\theta^\lambda\right)\eqno(59a)$$
$$\delta^{(1)} G_{\mu\nu}^\lambda = -\theta^\rho \partial_\rho G_{\mu\nu}^\lambda - \left(G_{\mu\rho}^\lambda \partial_\nu  + G_{\nu\rho}^\lambda \partial_\mu \right)
\theta^\rho + G_{\mu\nu}^\rho \partial_\rho\theta^\lambda\eqno(59b)$$
and
$$\delta^{(2)} h^{\mu\nu} = 0 = \delta^{(2)} G_{\mu\nu}^\lambda\,.\eqno(60a,b)$$
Eqs. (57, 59, 60) all can be generalized to $d > 2$ dimensions; this is the diffeomorphism transformation of eq. (5).

Together eqs. (56b) and (58) lead to
$$\hspace{-3cm}\overline{\delta}^{\,(1)} h^{\mu\nu} = -\left(\epsilon^{\mu\rho} h^{\nu\sigma} + \epsilon^{\nu\rho}h^{\mu\sigma}\right)\xi_{\rho\sigma}\eqno(61a)$$
and
$$\overline{\delta}^{\,(1)} G_{\mu\nu}^\lambda = \epsilon^{\lambda\rho}\left[\xi_{\mu\rho} G_{\sigma\nu}^\sigma +
\xi_{\nu\rho}G_{\sigma\mu}^\sigma - \xi_{\mu\sigma}G_{\rho\nu}^\sigma - \xi_{\nu\sigma}G_{\rho\mu}^\sigma  \right)\nonumber$$
which is equivalent to
$$\hspace{-1cm}= - \epsilon^{\rho\sigma}\left(G_{\mu\rho}^\lambda \xi_{\nu\sigma} + G_{\nu\rho}^\lambda \xi_{\mu\sigma}\right).\eqno(61b)$$
These in turn result in
$$\overline{\delta}^{\,(2)}h^{\mu\nu} = \overline{\delta}^{\,(2)} G_{\mu\nu}^\lambda = 0.\eqno(62a,b)$$
Together, eqs. (58, 61, 62) are the form of the gauge transformation that leaves the action of eq. (2) invariant when $d = 2$ that is derived in refs. [12, 14] from the
first class constraints associated with this action.

Generalizing eq. (58) to $d > 2$ dimensions by taking
$$\overline{\delta}^{\,(0)}h^{\mu\nu} = 0 \qquad\overline{\delta}^{\,(0)}G_{\mu\nu}^\lambda = -\epsilon^{\lambda\tau\rho_{1}\ldots\rho_{d-2} }
\xi_{\mu\nu\rho_{1}\ldots\rho_{d-2},\tau}\eqno(63a,b)$$
does not lead to consistent expressions for
$\overline{\delta}^{\,(2)}h^{\mu\nu}$ and $\overline{\delta}^{\,(2)}G_{\mu\nu}^\lambda$.  This indicates that while there are two invariances associated with the
action of eq. (2) when $d = 2$ (one of which is a ``gauge'' invariance in that it is generated by the first class constraints in the model), the only invariance present when
$d \neq 2$ is a diffeomorphism invariance.

If we follow the HTZ approach to determining the generator of a gauge transformation, then we consider an extended action
$$S_E = \int d^{d-1} x\, dt\left[ \pi_A \frac{\partial}{\partial t} \phi^A - {\cal{H}}_c\left(\phi^A, \pi_A\right) - U^{\alpha_{i}} \gamma_{\alpha{_{i}}} \left(\phi^A, \pi_A\right)\right]\eqno(64)$$
where $\phi^A$ is a field with conjugate momentum $\pi_A$, $\gamma_{\alpha{_{i}}}$ is a first class constraint of the $i^{th}$ generation and $U^{\alpha_{i}}$ is a
Lagrange multiplier.  (Second class constraints have been eliminated and all brackets are DBs.)

Variation of any function $F$ of $\phi^A$, $\pi_A$ is given by $\delta F = \left\lbrace F, G \right\rbrace$ where the generator $G$ is
$G = \int d^{d-1} y\lambda^{\alpha_{i}}\gamma_{\alpha{_{i}}}$ with $\lambda^{\alpha_{i}}$ being a gauge parameter.  This leads to (upon dropping a surface term)
$$\delta S_E = \int d^{d-1} x\, dt \left[ \frac{\partial\phi^A}{\partial t} \left\lbrace \pi_A, G\right\rbrace^* -
\frac{\partial \pi_A}{\partial t} \left\lbrace \phi^A, G\right\rbrace^* - \left\lbrace \phi^A, G\right\rbrace^* \frac{\delta {\cal{H}}_c}{\delta\phi^A}\right.\eqno(65)$$
$$\left. -\left\lbrace \pi^A, G\right\rbrace^* \frac{\delta {\cal{H}}_c}{\delta\pi_A} - \delta U^{\alpha_{i}} \gamma_{\alpha{_{i}}} - U^{\alpha_{i}} \left\lbrace \gamma_{\alpha_{i}}, G
\right\rbrace^*\right].\nonumber$$
If, as is the case of the EH action (see eq. (36)), the second class constraints are such that the DB and PB are identical in eq. (65), we find that
$$\delta S_E = \int d^{d-1} x\,dt\Biggl[ - \frac{\partial\phi^A}{\partial t} \frac{\delta G}{\delta \phi^A} - \frac{\partial\pi_A}{\partial t} \frac{\delta G}{\delta \pi_A} -
\left\lbrace {\cal{H}}_c, G \right\rbrace\nonumber$$
$$ - \delta U^{\alpha_{i}} \gamma_{\alpha{_{i}}} - U^{\alpha_{i}}\left\lbrace \gamma_{\alpha{_{i}}} , G\right\rbrace \Biggr].\eqno(66)$$
Since
$$ \frac{d\,G}{dt} = \int\left(\gamma_{\alpha{_{i}}} \frac{\partial\lambda^{\alpha{_i}}}{\partial t} + \frac{\partial\phi^A}{\partial t}  \frac{\delta G}{\delta \phi^A} +
 \frac{\partial\pi_A}{\partial t}  \frac{\delta G}{\delta \pi_A} + \gamma_{\alpha{_{i}}} \dot{U}^{\beta_{j}} \frac{\delta\lambda^{\alpha{_i}}}{\delta U^{\beta_{j}}} + \ldots\right)
 d^{d-1}y\nonumber$$
$$\hspace{-2cm}\equiv \int\left(\gamma_{\alpha{_{i}}} \frac{D\lambda^{\alpha{_i}}}{D t} + \frac{\partial\phi^A}{\partial t}  \frac{\delta G}{\delta \phi^A} +
 \frac{\partial\pi_A}{\partial t}  \frac{\delta G}{\delta \pi_A}\right)d^{d-1}y \eqno(67)$$
and $\left\lbrace U^{\alpha_{i}}, G\right\rbrace = 0$ we find that
$$\delta S_E = \int d^{d-1} x\,dt \left[ \frac{D\lambda^{\alpha{_i}}}{Dt} \gamma_{\alpha{_i}} + \left\lbrace G, {\cal{H}}_c + U^{\alpha_{i}}\gamma_{\alpha{_i}}\right\rbrace
-\delta U^{\alpha{_i}}\gamma_{\alpha{_i}}\right].\eqno(68)$$

Working in the gauge in which $U^{\alpha{_i}} = 0 = \delta U^{\alpha{_i}} (i \geq 2), \;\;\delta S_E = 0$ if
$$ \int d^{d-1} x \left[ \frac{D\lambda^{\alpha{_i}}}{Dt} \gamma_{\alpha{_i}} + \left\lbrace G, {\cal{H}}_c + U^{\alpha_{1}}\gamma_{\alpha{_1}}\right\rbrace
-\delta U^{\alpha{_1}}\gamma_{\alpha{_1}}\right] = 0.\eqno(69)$$
This condition can be used to find the gauge parameters $\lambda^{\alpha_{i}}$ that ensure that the ``total'' action
$$ S_T = \int d^{d-1} x\,dt \left(\pi_A \frac{\partial\phi^A}{\partial t} - {\cal{H}}_c\left(\phi^A , \pi_A\right) - U^{\alpha_{1}} \gamma_{\alpha_{1}}
\left(\phi^A, \pi_A\right)\right) \eqno(70)$$
is left invariant.  An invariance of the total action is an invariance of the initial action $S = \int d^{d-1} x\,dt \;\cal{L}$ [33].

The first and second order forms of the EH action are distinct when $d = 2$ [21, 22].  The canonical structure of the second order form appears in ref. [39]; the first order
form is discussed in refs. [12-17].

The action for $d = 2$ that follows from eq. (2) can be written
$$\hspace{-.2cm}S_2 = \int d^2x \left[ \left(-G_{00}^0 h_{,0} - 2G_{01}^0 h_{,0}^1 - G_{11}^0 h_{,0}^{11}\right) + \left( -G_{00}^1 \right) \left(h_{,1} + 2h G_{01}^0\right.\right. \eqno(71)$$
$$\left. + 2h^1 G_{11}^0\right) + \left(-2 G^1_{01}\right)\left(h_{,1}^1 - h G_{00}^0 + h^{11} G_{11}^0\right) + \left(-G_{11}^1\right)
\left(h_{,1}^{11} - 2h^1 G_{00}^0\right.\nonumber$$
$$\hspace{-6cm}\left.\left. - 2h^{11}G_{01}^0\right) \right]\nonumber$$
where $h = h^{00}$, $h^1 = h^{01}$.  We identify $\left(-G_{00}^0,\; -2G_{01}^0,\;-G_{11}^0\right)$ with the momenta $(\pi, \pi_1,\pi_{11})$ associated with
$(h, h^1, h^{11})$ respectively.  If $\zeta^1 = G_{00}^1, \zeta = 2G_{01}^1$, and $\zeta_1 = G_{11}^1$, then the Hamiltonian that follows
from eq. (71) is
$$H = \int dx\left[\zeta^1 \phi_1 + \zeta\phi + \zeta_1\phi^1\right]\eqno(72)$$
where
$$\phi_1 = h_{,1} - h\pi_1 - 2h^1\pi_{11},\quad\phi = h_{,1}^1 + h\pi - h^{11}\pi_{11}\quad\phi^1 = h_{,1}^{11} + 2h^1\pi +
h^{11}\pi_1\,.\eqno(73a,b,c)$$
The momenta $\Pi_1,\Pi,\Pi^1$ associated with $\zeta^1 , \zeta, \zeta_1$ respectively all vanish. These primary first class constraints lead to the secondary first
class constraints $\phi_1, \phi, \phi^1$ respectively.  There are no tertiary constraints since
$$\left\lbrace \phi_1, \phi^1\right\rbrace = 2\phi,\quad \left\lbrace \phi, \phi^1\right\rbrace = \phi^1 ,\quad  \left\lbrace \phi_1, \phi\right\rbrace = \phi_1\,.\eqno(74a,b,c)$$
Upon making the identification
$$\left(\gamma_{1_{1}},\, \gamma_{2_{1}},\,\gamma_{3_{1}},\,\gamma_{1_{2}},\,\gamma_{2_{2}},\,\gamma_{3_{2}}\right) =
\left(\Pi_1,\,\Pi,\,\Pi^1;\; \phi_1,\,\phi,\,\phi^1\right)\eqno(75)$$
then eq. (69) leads to
$$\lambda^{1_{1}} = \dot{\lambda}^{1_{2}} + \zeta\lambda^{1_{2}} - \zeta^1\lambda^{2_{2}}\eqno(76a)$$
$$\lambda^{2_{1}} = \dot{\lambda}^{2_{2}} + 2\zeta_1\lambda^{1_{2}} - 2\zeta^1\lambda^{3_{2}}\eqno(76b)$$
$$\lambda^{3_{1}} = \dot{\lambda}^{3_{2}} + \zeta_1\lambda^{2_{2}} - \zeta\lambda^{3_{2}}\,.\eqno(76c)$$
If now $\xi_{01} = \xi_{10} = -\frac{1}{2}\,\lambda^{2_{2}}$, $\xi_{11} = -\lambda^{3_{2}}$, $\xi_{00} = -\lambda^{1_{2}}$ then the generator
$G$ of the gauge transformations is seen to generate the transformation of eqs. (58, 61, 62).  This generator was obtained in ref. [12] by using the method of Castellani [29].

We reserve the term ``gauge transformation'' for a transformation generated by the first class constraints which leaves the action invariant in form. This does not preclude the existence of  transformations that leave the action invariant that are not a consequence of the existence of first class constraints.  In this sense, the action of eq. (2) when
$d = 2$ is invariant under a gauge transformation defined by eqs. (58, 61, 62) while the diffeomorphism transformation of eqs. (57, 59, 60), even though it is an invariance of the
action, is not referred to as a ``gauge transformation''. (It appears that the general belief is that any local transformation which leaves the action
invariant is a result of the presence of first class constraints; we see that this is not always the case.)
The fact that there might be a number of invariances associated with a model that are not what we call ``gauge invariances'' does not
increase the number of restrictions on the number of degrees of freedom present beyond those following from constraints that arise in the course of applying the Dirac constraint formalism;
for example, the presence of a diffeomorphism invariance in $S_2$ of eq. (2) does not reduce the number of degrees of freedom in $S_2$ as diffeomorphism invariance is not a consequence of the
first class constraints.  Indeed, the presence of diffeomorphism invariance in this model is of no consequence in the quantization of this model [17]; one need only consider the
invariance under the transformation of eqs. (58, 61, 62) when defining the path integral.  That is, the invariance associated with the diffeomorphism invariance does not require gauge
fixing and does not generate ghost fields - one need only consider the invariances of eqs. (58, 61, 62) when applying the Faddeev-Popov quantization procedure (or its extension [62])
associated with the path integral.

When $d > 2$, the generator of the gauge transformation is of the form
$$G = \int \left( a I\!\!P + a^i I\!\!P_i + b\chi + b^i\chi_i + c\tau + c^i\tau_i\right)d^{d-1}y \,,\eqno(77)$$
where eq. (70) is used to determine the coefficients $(a, a^i, b, b^i)$ in terms of $(c, c^i)$.  In fact, to find the variations $\delta h$, $\delta h^i$ and $\delta H^{ij}$ under
a gauge transformation, the coefficients $(a, a^i)$ are not required.  Furthermore, to obtain $(b, b^i)$ in terms of $(c, c^i)$, one need only ensure that eq. (70) is
satisfied by these terms linear in $(\tau , \tau_i)$; terms linear in $(\chi , \chi_i)$ fix the coefficients $(a, a^i)$ while terms linear in $(I\!\!P, I\!\!P_i)$ fix the
variations $(\delta U^1 ,\delta U^{1_{i}})$ of the Lagrange multiplier coefficients associated with the primary first class constraints $(I\!\!P,I\!\!P_i)$.

The Hamiltonian of eq.(45) and the secondary first class constraints of eqs. (16, 17) have the PBs
$$\left\lbrace\chi, {\cal{H}}_c\right\rbrace = \frac{1}{h} \left(\tau + h^i\tau_i\right) + \ldots\eqno(78)$$
$$\hspace{-1.7cm}\left\lbrace\chi_i , {\cal{H}}_c\right\rbrace = \tau_i + \ldots\eqno(79)$$
where only the terms dependent on $\tau$, $\tau_i$ have been displayed.  Furthermore, it follows from eqs. (45, 48-50) that
$$\hspace{-5.5cm}\left\lbrace \int dy \left(c\tau + c^i\tau_i \right), \int \,dx\,{\cal{H}}_c\right\rbrace\nonumber$$
$$\hspace{-2.3cm}= \int dx \Biggl[- \frac{(ch)_{,i}}{h^2} \,H^{ij} \tau_j + \left(\frac{(hc)_{,i}h^i}{h^2} - \frac{ch_{,i}^i}{h}\right)\tau\eqno(80)$$
$$+ \frac{(hc^i)_{,i}}{h^2} \,\tau + \left(\frac{c_{,j}^ih^j}{h}\tau_i - \frac{c^ih_{,i}^j}{h}\tau_j +
\frac{c^ih^jh_{,i}}{h^2}\,\tau_j\right) + \ldots\Biggr]\nonumber$$
where again only terms dependent on $\tau$, $\tau_i$ are given explicitly.

From eqs. (78-80) it follows that eq. (70) is satisfied by those terms linear in $(\tau, \tau_i)$ provided
$$ \frac{\partial c}{\partial t} + \frac{b}{h} + \frac{(hc)_{,i}h^i}{h^2} - \frac{ch_{,i}^i}{h} + \frac{(hc^i)_{,i}}{h^2} = 0\eqno(81)$$
and
$$ \frac{\partial c^i}{\partial t} + \frac{bh^i}{h} + b^i - \frac{(ch)_{,j}}{h^2}\, H^{ij} + \frac{c_{,j}^ih^j}{h} - \frac{c^jh_{,j}^i}{h} + \frac{c^jh^ih_{,j}}{h^2} = 0\,.\eqno(82)$$
Eqs. (81, 82) fix $b$ and $b^i$ in terms of $c(x,t)$, $c^i(x,t)$ to be
$$\hspace{-1.6cm}b = -hc_{,t} - \frac{(hc)_{,j}h^j}{h} + ch_{,j}^j - \frac{\left(hc^j\right)_{,j}}{h}\eqno(83)$$
$$b^i = -c^i_{,t} + h^ic_{,t} + \frac{h^i(hc)_{,j}h^j}{h^2} - \frac{h^ich^j_{,j}}{h}+ \frac{h^i\left(hc^j\right)_{,j}}{h^2}\nonumber$$
$$+ \frac{(ch)_{,j}}{h^2} \,H^{ij} - \frac{c^i_{,j}h^j}{h} + \frac{c^jh^i_{,j}}{h} - \frac{c^jh^ih_{,j}}{h^2}\,.\eqno(84)$$

Using eqs. (83, 84) the generator of eq. (77) leads to the gauge transformation of $h$, $h^i$ and $H^{ij}$ (and all other fields).  We find that
$$ \hspace{-2cm}\delta h = \left\lbrace h, G\right\rbrace = \left\lbrace h, \int d^{d-1} y(b\chi)\right\rbrace\nonumber$$
$$= -h^2 c_{,t} - (hc)_{,j} h^j + chh_{,j}^i - (hc^j)_{,j}\eqno(85)$$
$$\hspace{-1.5cm} \delta h^i = \left\lbrace h^i, G\right\rbrace = \left\lbrace h^i, \int d^{d-1} y(b^j\chi_j)\right\rbrace\eqno(86)$$
$$= hc_{,t}^i - h^ihc_{,t} + h^i ch_{,j}^j - \frac{h^i(hc^j){_{,j}}}{h}\nonumber$$
$$-(c h)_{,j}h^{ij} + c^i_{,j}h^j - c^jh_{,j}^i + \frac{c^jh^ih_{,j}}{h}\nonumber$$
$$\hspace{-2.5cm}\delta H^{ij} = \left\lbrace H^{ij}, G \right\rbrace = \left\lbrace H^{ij}, \int d^{d-1}y(c\tau + c^k\tau_k)\right\rbrace\eqno(87)$$
$$= c(H^{ij} H^{k\ell} - H^{ik}H^{j\ell})\Pi_{k\ell} - (H^{ik}c^j + H^{jk}c^i)_{,k}\nonumber$$
$$ + (H^{ij}c^k)_{,k} + c^k_{,k}H^{ij}.\nonumber$$
From eq. (5b) we find that
$$\hspace{-1cm}\delta h = 2h\theta_{,t} + 2h^i\theta_{,i} - (h\theta)_{,t} - (h\theta^i)_{,i}\eqno(88)$$
$$\delta h^i = h\theta_{,t}^i + h^j\theta_{,j}^i - h^i_{,t}\theta + h^{ij}\theta_{,j} - (h^i\theta^j)_{,j}\eqno(89)$$
$$\hspace{-3cm}\delta h^{ij} = \delta \left( \frac{H^{ij} + h^ih^j}{h}\right)\eqno(90)$$
$$= h^i\theta^j_{,t} + h^j\theta^i_{,t} + h^{ik}\theta_{,k}^j + h^{jk}\theta_{,k}^i\nonumber$$
$$- (h^{ij}\theta )_{,t} - (h^{ij}\theta^k)_{,k}.\nonumber$$
Eqs. (85) and (86) can be reconciled by making the field dependent redefinition $\theta = -hc$, $\theta^i = c^i - h^ic$, but the presence of $\Pi_{k\ell}$ in eq. (87)
prevents us from reconciling eqs. (87) and (90) in the same way.  Consequently the invariance of the first order EH action uncovered by our application of the HTZ
formalism is not diffeomorphism invariance. We have not as yet explicitly examined how the affine connections transform.

In a superficially similar situation, the gauge invariance of the ADM action considered in refs. [29, 36, 37] is only consistent with diffeomorphism if there is a
field dependent gauge parameter, while the gauge invariance that follows from the second order EH action, when using the metric $g_{\mu\nu}$ as the configuration space variable, is
the diffeomorphism invariance of eq. (5b) [8,9].

All this prompts us to reflect on the way in which gauge invariance is related to the constraints in a system.  We begin by noting that the change of variables that has taken us from
the EH action of eq. (1) written in terms of the configuration space metric $g_{\mu\nu}$ and the affine connection $\Gamma_{\mu\nu}^\lambda$ to where it is written in eq. (13) in terms
of the variables $h$, $h^i$, $H^{ij}$ etc. of course does not alter the equations of motion derived by applying the principle of least action to the Lagrangian
form of the action; using either eq. (1) or (13) will result in the Einstein equations of motion.  If $q_i$ and $Q_i$ denote the set of old and new configuration space variables
respectively, then if there were no constraints one could pass from the Lagrangian to the Hamiltonian formalism using either $L(q_i, \dot{q}_i)$ or $L(Q_i, \dot{Q}_i)$ (The transformation
from $q_i$ to $Q_i$ is invertible.).  The phase space variables $(q_i, p_i = \partial L/\partial\dot{q}_i)$ and $(Q_i, I\!\!P_i = \partial L/\partial\dot{Q}_i)$ would then be
related by a canonical transformation, and the Hamilton equations of motion derived from using either $H(q_i, p_i)$ or $H(Q_i,I\!\!P_i)$ would be equivalent to each other and
to the Lagrange equations of motion following from either $L(q_i,\dot{q}_i)$ or $L(Q_i,\dot{Q}_i)$.

When there are constraints in a theory as in the EH action, more care must be taken when changing variables.  One can change variables in configuration space from $q_i$ to $Q_i$ and
obtain equivalent equations of motion from either $L(q_i,\dot{q}_i)$ or $L(Q_i,\dot{Q}_i)$ as in the case when there are no constraints.  However, when passing from the
Lagrangian to the Hamiltonian formalism, the canonical variables $(q_i,p_i)$ and $(Q_i,I\!\!P_i)$ may not be related by a canonical transformation when there are constraints
in the theory.  This is explicitly demonstrated in refs. [8,9] in the context of passing from using the configuration space variables $g_{\mu\nu}$ to $(g_{ij},N,N_i)$ of the
ADM formalism when treating the second order form of the EH action.  In these references, the phase space variables $(g_{\mu\nu}, p^{\mu\nu})$ derived from $L(g_{\mu\nu})$ are shown
to not be canonical transforms of the phase space variables $(g_{ij}, N, N_i, \Pi^{ij}, \Pi , \Pi^i)$ derived from $L(g_{ij}, N, N_i)$.  This is despite the fact
that the Hamilton equations of motion derived from $H(g_{\mu\nu}, p^{\mu\nu})$ and $H(g_{ij}, N, N_i, \Pi^{ij}, \Pi , \Pi^i)$ are both equivalent to the Einstein
field equations.  Nevertheless, there is a significant difference between the actions in phase space written in terms of these two sets of variables that is pointed out in
refs.[8,9]; the action in terms of $(g_{\mu\nu}, p^{\mu\nu})$ can be used to derive the diffeomorphism gauge invariance of the action while the action in terms of
$(g_{ij}, N, N_i, \Pi^{ij}, \Pi , \Pi^i)$ can only be used to derive a diffeomorphism gauge invariance with field dependent gauge functions.  (The group properties of gauge
transformations when there are field dependent gauge functions is discussed in refs. [67, 37].)

It is apparent though that when starting from the first order action of eq. (1) in terms of $g_{\mu\nu}, \Gamma_{\mu\nu}^\lambda$ the phase space position and momentum
variables are related by a canonical transformation to those derived from eq. (13).  By construction, the generator $G$ given in eq. (77) will provide a gauge invariance of the
action defined in terms of the phase space variables being used.  However this invariance may not be unique.  The tertiary constraints $(\tau, \tau_i)$ appearing in eq. (77) could be
supplemented by a function of the secondary constraints so that instead of $(\tau, \tau_i)$ appearing in eq. (77), one could have
$$\tilde{\tau} = \tau + X\chi + X^i \chi_i\eqno(91)$$
$$\tilde{\tau}_i = \tau_i + Y_i \chi + Y_i^j \chi_j  \eqno(92)$$
where $X, X^i, Y_i, Y_,^j$ are arbitrary functions of the dynamical variables such as $h,\Pi$ etc.  Indeed, in determining the tertiary constraints from examining
$\left\lbrace \chi , {\cal{H}}_c\right\rbrace$ and $\left\lbrace \chi_i , {\cal{H}}_c\right\rbrace$ with ${\cal{H}}_c$ given by eq. (45), it is not $\tau$ and $\tau_i$ that immediately appear,
but rather expressions of the form of eqs. (91, 92).  With $\tilde{\tau}$ and $\tilde{\tau}_i$ now appearing in eq. (77), the solution for $b$ and $b^i$ will no longer
be given by eqs. (83, 84) and so the gauge transformation generated by $G$ will be altered.  This is because the analogue of eq. (80) with $(\tilde{\tau}, \tilde{\tau}_i)$ appearing
in place of $(\tau, \tau_i)$ does not follow if $X$, $X^i$, $Y_i$, $Y_i^{\;j}$ are arbitrary and hence $(b, b^i)$ are dependent on the form of the tertiary constraints
used in eq. (77).  Thus the invariances of the original action which follow from the generator $G$ are dependent on the ansatz used initially for $G$; it is not apparent
which ansatz leads to a diffeomorphism.

The ambiguity present in the HTZ formalism that has been noted here is likely to be absent [60] in the approach of Castellani [29].  To see this, we first will sketch the way
in which the gauge generator $G$ can be derived using the methods of ref. [29].

If a system has canonical variables $(q_i, p_i)$, and a gauge generator $G$, then $(q_i(t), p_i(t))$ and $(q_i(t) + \alpha_i(t), p_i(t) + \beta_i(t))$ would both be solutions of the
equations of motion if
$$\alpha_i = \left\lbrace q_i, G\right\rbrace = \frac{\partial G}{\partial p_i}\quad {\mathrm{and}}\quad
\beta_i = \left\lbrace p_i, G\right\rbrace = -\frac{\partial G}{\partial q_i} \eqno(93)$$
so that by the weak equation [23, 33] $\frac{d}{\partial t} A(q_i,p_i,t) \approx \left\lbrace A(q_i,p_i,t), H_T\right\rbrace + \frac{\partial A(q_i,p_i,t)}{\partial t }$
$$\dot{\alpha}_1  \approx \left\lbrace \frac{\partial G_i}{\partial p_i}, H_T\right\rbrace + \frac{\partial^2 G}{\partial t\partial p_i} \quad \mathrm{and}\quad
\dot{\beta}_1  \approx -\left\lbrace \frac{\partial G}{\partial q_i}, H_T\right\rbrace - \frac{\partial^2 G}{\partial t\partial q_i}. \eqno(94)$$
Furthermore, the equations of motion themselves yield
$$\dot{q}_i + \dot{\alpha}_1 \approx \frac{\partial}{\partial p_i} H_T (q_i + \alpha_i, p_i + \beta_i) \quad \mathrm{and} \quad
\dot{p}_i + \dot{\beta}_1 \approx -\frac{\partial}{\partial q_i} H_T (q_i + \alpha_i, p_i + \beta_i)\eqno(95)$$
or to lowest order in $\alpha_i$ and $\beta_i$
$$\dot{\alpha}_i \approx \frac{\partial}{\partial p_i} \left(\frac{\partial H_T}{\partial q_j} \alpha_j + \frac{\partial H_T}{\partial p_j} \beta_j\right) \quad \mathrm{and}
\quad \dot{\beta}_i \approx -\frac{\partial}{\partial q_i} \left(\frac{\partial H_T}{\partial q_j} \alpha_j + \frac{\partial H_T}{\partial p_j} \beta_j\right).\eqno(96)$$
(The weak equality is one which holds if the primary constraints vanish.)

If now there are three generations of constraints, we take
$$G = \epsilon (t) G_0 + \dot{\epsilon}(t) G_1 + \ddot{\epsilon} (t) G_2.\nonumber$$

Upon equating our two expressions (94, 96) for $\dot{\alpha}_i,\dot{\beta}_i$ and eliminating $\alpha_i$ and $\beta_i$ using eq. (93) we derive the ``master equation''
$$\left[\epsilon\left\lbrace G_0,H_T\right\rbrace + \dot{\epsilon}(t) \left(G_0 + \left\lbrace G_1,H_T\right\rbrace\right) + \ddot{\epsilon}(t)
\left(G_1 + \left\lbrace G_2,H_T\right\rbrace\right) + {\stackrel{\dots}{\epsilon}}(t)G_2\right] \approx 0.\eqno(97)$$

We now identify the primary constraints $I\!\!P_A = (I\!\!P, I\!\!P_i)$, secondary constraints $\chi_A = (\chi, \chi_i)$ and tertiary constraints $\tau_A = (\tau, \tau_i)$.  With
these sets of constraints and the canonical Hamiltonian of eq. (45), we have equations of the form
$$ \left\lbrace I\!\!P_A, H_T\right\rbrace = \chi_A,\quad \left\lbrace \chi_A, H_T\right\rbrace = V_{AB} \tau_B + \overline{V}_{AB}\chi_B,\quad \left\lbrace \tau_A, H_T\right\rbrace =
W_{AB}\tau_B + \overline{W}_{AB}\chi_B\eqno(98)$$
so that from the master equation (97)
$$\hspace{-7.3cm}G_2 \approx 0  \Rightarrow G_2 = I\!\!P_A\eqno(98)$$
$$\hspace{-3cm}G_1 + \left\lbrace G_2, H_T\right\rbrace \approx 0 \Rightarrow G_1 = -\chi_A+\lambda_{AB} I\!\!P_B\eqno(99)$$
$$\hspace{1cm}\mathrm{(for\;some\;\lambda_{AB})}\nonumber$$
$$G_0 + \left\lbrace G_1, H_T\right\rbrace \approx 0 \Rightarrow G_0 = \kappa_{AB}I\!\!P_B + (V_{AB}\tau_B + \overline{V}_{AB}\chi_B)\eqno(100)$$
$$- \left\lbrace \lambda_{AB}, H_T\right\rbrace I\!\!P_B - \lambda_{AB}\chi_B\nonumber$$
$$\mathrm{(for\;some\;\kappa_{AB})}\nonumber$$
and
$$\left\lbrace G_0, H_T\right\rbrace \approx 0 \Rightarrow \kappa_{AB}\chi_B + \left\lbrace \kappa_{AB},H_T\right\rbrace I\!\!P_B + V_{AB}
(W_{BC}\tau_C + \overline{W}_{BC}\chi_C)\nonumber$$
$$\hspace{-4.5cm}+\left\lbrace V_{AB}, H_T\right\rbrace \tau_B\nonumber$$
$$+\left\lbrace \overline{V}_{AB}, H_T\right\rbrace \chi_B + \overline{V}_{AB}\left[V_{BC} \tau_C + \overline{V}_{BC}\chi_C\right]\nonumber$$
$$-2\left\lbrace \lambda_{AB}, H_T\right\rbrace ]\chi_B - \left\lbrace\left\lbrace \lambda_{AB}, H_T\right\rbrace, H_T\right\rbrace I\!\!P_B\nonumber$$
$$-\lambda_{AB} \left[V_{BC} \tau_C + \overline{V}_{BC}\chi_C\right] \approx \rho_{AB}I\!\!P_B.\eqno(101)$$
$$\mathrm{(for\;some\;\rho_{AB})}\nonumber$$
(The quantities $\rho$, $\kappa$ and $\lambda$ may be non-local.)
This last equation is satisfied if the coefficients of $I\!\!P_A$, $\chi_A$ and $\tau_A$ all vanish, so that
$$\hspace{-3.5cm}\left\lbrace \kappa_{AB}, H_T\right\rbrace - \left\lbrace \left\lbrace \lambda_{AB}, H_T\right\rbrace, H_T\right\rbrace = \rho_{AB}\eqno(102)$$
$$\kappa_{AB} + V_{AP} \overline{W}_{PB} + \left\lbrace \overline{V}_{AB}, H_T\right\rbrace + \overline{V}_{AP}\overline{V}_{PB} - 2
\left\lbrace \lambda_{AB}, H_T\right\rbrace\eqno(103)$$
$$- \lambda_{AP}\overline{V}_{PB} = 0\nonumber$$
$$\hspace{-2cm}V_{AP}W_{PB} +  \left\lbrace V_{AB}, H_T\right\rbrace + \overline{V}_{AP}\overline{V}_{PB} - \lambda_{AP} V_{PB}  = 0.\eqno(104)$$
These equations can be solved for $\lambda_{AB}$, $\kappa_{AB}$ and $\rho_{AB}$ using eqs. (104), (103) and (102) in turn leading to a unique gauge generator $G$.  An
explicit calculation is quite formidable (especially on account of the complicated structures of $\overline{V}_{AB}$) and is currently being considered.  However, the
procedure of ref. (29) outlined here for obtaining the generator of a gauge transformation appears to be unambiguous, once the primary constraints are found, in contrast to the
HTZ method [30, 31, 25] discussed above.  In particular, it is insensitive to how one identifies $\tau_A$ in eqs. (91, 92), though it is dependent on the choice of primary constraints [59].

\section{Discussion}

From the outset, we have applied in a fully consistent way the Dirac constraint formalism to the $d$ dimensional EH action.  This has led to primary and secondary second
class constraints as well as primary, secondary and tertiary first class constraints, leaving $d(d-3)$ degrees of freedom in phase space.  The gauge transformations which leave the first order EH
action invariant in $d$ dimensions that is implied by the first class constraints do not
appear to coincide with the diffeomorphism transformation when the HTZ formalism is used.

It would be interesting to analyze the implications of having not only primary and secondary, but also tertiary first class constraints (and all their attendant gauge
conditions) on the quantization of the first order EH action of eq. (2).  The quantization of this action was considered in ref. [40] using the Faddeev-Popov-Feynman-deWitt-Mandelstam
quantization procedure with the diffeomorphism of eq. (5) as the gauge invariance of the theory, although explicit calculations do not appear to have been performed using the first order form of the action. (See however ref. [41].)  The first
order form has an advantage over the second order form in that its interaction is only cubic as opposed to being non-polynomial; even in Yang-Mills theory the first order form has
calculational advantages [42].

If one were to use the path integral to quantize this model, the non-trivial second class constraints $\Theta_I^a = \left(\overline{I\!\!P}_j^i, I\!\!P_i^{jk}, \Theta_j^i, \Theta_i^{jk}\right)$
must be taken into account in the measure of the functional integration.  This is because a factor of $\det^{1/2}\left\lbrace \Theta_I^a, \Theta_J^b\right\rbrace$ occurs in this measure
[57], and from eqs. (23, 24), this factor is non-trivial. It is not clear how this factor would be generated if one were to apply the Faddeev-Popov procedure (or
its extension [62]) for using the path integral to quantize
a gauge theory.  A more elaborate approach [63] is likely required.

The first order action for general relativity when expressed in terms of the spin connection and tetrad (the Einstein Cartan (EC) action) is not equivalent to the EH action in
that the tetrad cannot be uniquely expressed in terms of the metric [43].  A canonical analysis of this EC action for $d = 3$ [44] and $d > 3$ [45] dimensions reveals
that its first class constraints generate translational and rotational transformations in the tangent space and cannot generate the diffeomorphism transformation.  Such
transformations have been found in refs. [64, 65, 66].  For the vierbein $e_\mu^a$ and the spin connection $\omega_{\mu ab}$ it was found that there is both the rotational
invariance
$$\delta_r\omega_{\mu ab} = -\partial_\mu r_{ab} - \left(\omega_{\mu a}^{\;\;\;\;c}r_{cb} - r_a^{\;\;c}\omega_{\mu cb}\right)\eqno(105a)$$;
$$\delta_r e_{\mu a} = r_{a}^{\;\;b}e_{\mu b}\eqno(105b)$$
and the translational invariance
$$\delta_t \omega_{\mu ab} = R_{\mu\lambda ab} e^{\lambda c}t_c \left( R_{\mu\nu ab} \equiv \partial_\mu \omega_{\nu ab} - \partial_\nu \omega_{\mu ab} + \omega_{\mu ac}
\omega_{\nu\;\;\;b}^{\;c} - \omega_{\nu ac} \omega_{\mu\;\;\;b}^{\;c}\right)\eqno(106a)$$
$$\delta_t e^{\mu a} = e^{\rho a} (e^{\mu c} t_c)_{,\rho} - \left[\partial_\rho e^{\mu a} + \omega_{\rho ab} e^{\mu b}\right](e^{\rho c}t_c).\eqno(106b)$$
The transformations of eqs. (105, 106) are in tangent space.  They are related to the usual diffeomorphism transformations
$$\delta e^{\mu a} = -e^{\mu a}_{\;\;,\lambda} \xi^\lambda + e^{\lambda a}\xi^\mu_{,\lambda}\eqno(107a)$$
$$\delta \omega^{\mu ab} = -\omega^{\mu ab}_{\;\;,\lambda} \xi^\lambda + \omega^{\lambda ab}\xi^\mu_{,\lambda}\eqno(107b)$$
by [66]
$$t_a = e_{\lambda a} \xi^\lambda\eqno(108a)$$
$$r_{ab} = \omega_{\lambda ab} \xi^\lambda\;.\eqno(108b)$$
All indications are [45] that the transformations of eqs. (105, 106) and not those of eqs. (107) are generated by the first class constraints arising from the EC action.  Once again,
as in the $2D$ first order EH action, not all invariances are generated by the first class constraints in the theory.  We do note though that the diffeomorphism transformation of
eq. (107) can be found using eqs. (54) [68].
This will likely affect the quantization of the EC action, since much like the case of the action $S_2$ of eq. (2) being quantized, the diffeomorphsim invariance that is present
is not to be associated with the presence of ghosts [17]. (In ref. [58], however, the translational invariance of the EC action was ignored and diffeomorphism invariance was in
fact used to generate ghost fields.)

\section{Appendix A: Canonical Analysis of the Spin Two Field in First Order Formalism}

We now apply the Dirac constraint formalism to the first order form of the spin two action as it differs in interesting ways from that of the EH action considered in the
body of the paper.  Various aspects of this problem have been discussed in [46-54].

In order to linearize the action of eq. (2) we replace it by
$$\overline{S}_d = \frac{1}{2} \int\, d^dx \left[h^{\mu\nu} G_{\mu\nu ,\lambda}^\lambda + \eta^{\mu\nu} \left(\frac{1}{d-1} G_{\lambda\nu}^{\lambda} G_{\sigma\nu}^\sigma -
G_{\sigma\mu}^\lambda G_{\lambda \nu}^\sigma\right)\right]\eqno(A1)$$
provided $d > 2$.  (The case of $d = 2$ will be dealt with below.)  Here we use the flat space metric $diag \,\eta^{\mu\nu} = (-, + \ldots +)$.

Expressing $G_{\mu\nu}^\lambda$ in terms of $h_{\mu\nu}$ using the equations of motion leads to
$$\overline{S}_d = \int d^dx \left[ h_{,\sigma}^{\mu\lambda} h_{\;\;\mu \,,\lambda}^\sigma - \frac{1}{2} h^{\mu\nu}_{\;,\lambda} h_{\mu\nu ,}^{\;\;\;\lambda} +
\frac{1}{2(d-2)} \,h_{\;\;\mu \,,\lambda}^\mu h_{\;\nu ,}^{\nu\;\;\lambda}\right]\eqno(A2)$$
which when $d = 4$ is the spin two action in refs. [55, 56].

If now we define
$$\pi = -G_{00}^0,\qquad\pi_i = - 2G_{0i}^0,\qquad\pi_{ij} = - G_{ij}^0\eqno(A3a-c)$$
and
$$\xi^k = G_{00}^k,\qquad \xi_j^i = 2G_{j0}^i = \overline{\zeta}^{\,i}_j + \frac{1}{d-1} t\delta_j^i,\qquad\xi_{jk}^i = G_{jk}^i\eqno(A4a-c)$$
where $\overline{\zeta}^{\,i}_i = 0$, then the canonical Hamiltonian density is
$$\hspace{-7cm}{\cal{H}}_c = \pi h_{,0} + \pi_ih_{,0}^i + \pi_{ij} h_{\,,0}^{ij} - {\cal{L}}\nonumber$$
$$ = \frac{2-d}{d-1} \left(\pi^2 - \frac{1}{4} \pi_i\pi_i\right) + \xi^k\left(\pi_k + h_{,k}\right) + \frac{t}{d-1} \left(-\pi_{ii} - \pi + h_{,i}^i\right)\nonumber$$
$$\hspace{-6cm}+ \overline{\zeta}^{\,i}_j \left(-\pi_{ij} + h_{,i}^j\right) - \frac{1}{4} \;\overline{\zeta}^{\,i}_j\,\overline{\zeta}^{\,j}_i\eqno(A5)$$
$$ + \xi_{jk}^i\left(h_{,i}^{jk} + \frac{1}{d-1}\,\delta_i^j\pi_k\right) +
\xi_{jk}^i\xi_{ik}^j - \frac{1}{d-1} \,\xi_{ik}^i\xi_{jk}^j\;.\nonumber$$
The momenta $I\!\!P_k$ and $I\!\!P$ conjugate to $\xi^k$ and $t$ vanish leading to the secondary constraints
$$\hspace{-1cm}\chi_k = h_{,k} + \pi_k\eqno(A6)$$
$$\chi = h_{,i}^i - \pi_{ii} - \pi\,.\eqno(A7)$$
The momenta conjugate to $\overline{\zeta}^{\,i}_j$ and $\xi^i_{jk}$ also vanish; these momenta and the equations of motion associated with these variables
obviously form a set of second class constraints.  Using their equations of motion, $\overline{\zeta}^i_j$ and $\xi^i_{jk}$ can then be eliminated from ${\cal{H}}_c$ in eq. (A5) to yield
$$\hspace{-4cm}{\cal{H}}_c = \frac{2-d}{d-1} \pi^2 + \frac{d-3}{4(d-2)} \,\pi_i\pi_i + \xi^k \left(\pi_k + h_{,k}\right)\eqno(A8)$$
$$+ \frac{t}{d-1}\left(-\pi_{ii} - \pi + h_{,i}^i\right)\nonumber$$
$$+\left(\pi_{ij}\pi_{ij} - \frac{1}{d-1} \pi_{ii}\pi_{jj} - 2 \pi_{ij} h_{,j}^i + \frac{2}{d-1}\, \pi_{kk} h_{,\ell}^\ell\right.\nonumber$$
$$\left. + \frac{d-2}{d-1} \,h_{,k}^k h_{,\ell}^\ell\right)\nonumber$$
$$-\left(\frac{1}{2(d-2)} h^{ii}_{\;,j}\,\pi_j + \frac{1}{2} h_{,i}^{jk}\,h^{ik}_{\;\;\;,j} + \frac{1}{4(d-2)} h^{mm}_{\;\;\;\;,j}\,h^{nn}_{\;\;\;\;,j}\right.\nonumber$$
$$\left. - \frac{1}{4} h^{mn}_{\;\;\;,j}h^{mn}_{\;\;\;,j}\right)\,.\nonumber$$
The secondary constraints of eqs. (A6), (A7) satisfy (with $H_c = \int d^{d-1}x{\cal{H}}_c$)
$$\hspace{-3cm}\left\lbrace H_c, \chi\right\rbrace = -\tau\eqno(A9)$$
$$\left\lbrace H_c, \chi_k\right\rbrace = -2\left(\frac{d-2}{d-1} \chi_{,k} + \tau_k\right) \eqno(A10)$$
so we have the tertiary constraints
$$\tau = h_{ij,ij} + \pi_{i,i}\eqno(A11)$$
$$\tau_k = \pi_{ii,k} - \pi_{ik,i}\;.\eqno(A12)$$
Since $\left\lbrace \tau,  H_c\right\rbrace = 0$, $\left\lbrace \tau_k,  H_c\right\rbrace = -\frac{1}{2} \tau_{,k}$ there are no fourth generation constraints.  All
constraints ($I\!\!P, I\!\!P_k, \chi, \chi_k, \tau, \tau_k$) have vanishing PB$s$ with each other and hence all are first class.  It can be shown using eq. (70) that generator of the
gauge transformation is given by
$$G = \int d^{d-1}x\left[\left(- (d-1)\ddot{\epsilon} + \frac{1}{4} (d-3) \dot{\epsilon}_{k,k}\right) I\!\!P - \frac{1}{4}\ddot{\epsilon}_k I\!\!P_k\right.\eqno(A13)$$
$$\left. - \left(\frac{1}{4} \epsilon_{k,k} + \dot{\epsilon}\right)\chi - \frac{1}{4} \dot{\epsilon}_k \chi_k + \epsilon\tau + \frac{1}{2} \epsilon_k\tau_k\right]\;.\nonumber$$
This leads to the spin two gauge transformation
$$\hspace{-1.3cm}\delta h^{\mu\nu} = \partial^\mu f^\nu + \partial^\nu f^\mu - \eta^{\mu\nu}\partial \cdot f\eqno(A14a)$$
$$ \delta G_{\mu\nu}^\lambda = - \partial^2_{\mu\nu} f^\lambda + \frac{1}{2} \left(\delta_\mu^\lambda \partial_\nu + \delta_\nu^\lambda \partial_\mu\right)\partial \cdot f\eqno(A14b)$$
which is a linearized version of eq. (5). Without tertiary constraints, the second derivates appearing in eq. (A14b) would not appear.

If $d = 2$, then one cannot solve for $G_{\mu\nu}^\lambda$ in terms of $h_{\mu\nu}$ using the action of eq. (A1).  However, if one were to set
$$G_{\mu\nu}^\lambda = \overline{G}_{\mu\nu}^{\,\lambda} + V^\lambda \eta_{\mu\nu}\;\;\;\left(\eta^{\mu\nu} \overline{G}_{\mu\nu}^{\,\lambda} \equiv 0\right)\eqno(A15)$$
then one can solve for $\overline{G}_{\mu\nu}^{\,\lambda}$ in terms of $h_{\mu\nu}$.  If this solution is substituted back into the action we find that
$$\overline{S}_2 = -\int d^2x h^{\mu\nu}_{\;\;,\lambda}\, \eta_{\mu\nu}V^\lambda\;,\eqno(A16)$$
showing the triviality of the theory when $d = 2$.  If we set $h = h^{00}$, $h^1 = h^{01}$, $\pi = -G_{00}^0$, $\pi_1 = -2 G_{01}^0$, $\pi_{11} = -G_{11}^0$,
$\xi = G_{00}^1$, $\xi_1 = 2G_{01}^1$, $\xi_{11} = G_{11}^1$, then $\overline{S}_2$ becomes
$$\overline{S}_2 = \int d^2x \left[ h_{,0}\pi + h_{,0}^1 \pi_1 + h^{11}_{\;,0} \pi_{11} - \xi\phi_1 - \xi_1\phi - \xi_{11}\phi^1\right]\eqno(A17)$$
so that
$$\phi_1 = h_{,1} + \pi_1\eqno(A18a)$$
$$\qquad\phi = h_{,1}^1 - \pi - \pi_{11}\eqno(A18b)$$
$$\phi^1 = h_{,1}^{11} + \pi_1\eqno(A18c)$$
are all first class constraints.  Any two of these constraints have a vanishing PB.

\section{Appendix B: Inclusion of Scalars}

We can supplement the action of eq. (2) with
$$S_\phi = \frac{1}{2} \int d^dx \,h^{\mu\nu} \left(\partial_\mu\phi\right)\left(\partial_\nu\phi\right)\,.\eqno(B1)$$
This does not alter the primary and secondary constraints of eqs. (20, 21).  From eq. (B1) though, the momentum associated with $\phi$ is
$$p = h\phi_{,0} + h^i\phi_{,i}\eqno(B2)$$
so that the Hamiltonian gets supplemented by
$${\cal{H}}_\phi = \frac{1}{2h} \left( p^2 - H^{ij} \phi_{,j}\phi_j\right) - \frac{ph^i\phi_{,i}}{h}\eqno(B3)$$
$$\hspace{-1.9cm} = \overline{{\cal{H}}}_\phi - \frac{ph^i\phi_{,i}}{h}\,.\nonumber$$
Since
$$\left\lbrace \chi, {\cal{H}}_\phi\right\rbrace = {\cal{H}}_\phi\eqno(B4)$$
$$\hspace{.45cm}\left\lbrace \chi_i, {\cal{H}}_\phi\right\rbrace = -p\phi_{,i}\eqno(B5)$$
the tertiary constraints of eqs. (46, 47) become
$$T_i = \tau_i - p\phi_{,i}\eqno(B6)$$
$$\hspace{.5cm}T = \tau + \frac{1}{h} \,\overline{{\cal{H}}}_{\phi}\;.\eqno(B7)$$
Since
$$\left\lbrace -p(x)\phi_{,i}(x),\;\;-p(y)\phi_{,j}(y)\right\rbrace = - \partial^x_j \delta(x-y) \left(-p(y)\phi_{,i}(y)\right)+ \partial_i^y \delta(x-y)\left(-p(x)\phi_{,j}(x)\right)\eqno(B8)$$

$$\left\lbrace \overline{{\cal{{H}}}}_\phi(x), \; \;\overline{{\cal{H}}}_\phi(y)\right\rbrace = \partial_i^x\delta(x-y) \frac{H^{ij}(y)}{h^2(y)} \left(-p(y)\phi_{,j}(y)\right)
-\partial_i^y \delta(x-y) \frac{H^{ij}(x)}{h^2(x)} \left(-p(x)\phi_{,j}(x)\right)\eqno(B9)$$

$$\hspace{.3cm}\left\lbrace \chi , \overline{\cal{{H}}}_\phi\right\rbrace = \overline{{\cal{H}}}_\phi\eqno(B10)$$
$$\left\lbrace \chi_i , \overline{{\cal{H}}}_\phi\right\rbrace = 0\eqno(B11)$$
we find that the form of the PBs of ($I\!\!P, I\!\!P_i, \chi, \chi_i, \tau, \tau_i$) with each other are of the same as the PBs of ($I\!\!P, I\!\!P_i, \chi, \chi_i, T, T_i$) with
each other.

\section{Acknowledgements}

The author would like to thank R. Ghalati for discussions on the algebra of tertiary constraints, and
N. Kiriushcheva and S. Kuzmin for discussions on the nature of gauge invariances in the action. The referee had a number
of constructive comments and suggestions.  R. Macleod was quite helpful.

\end{document}